\documentstyle[amssymb,12pt]{article}
\input{tcilatex}
\begin{document}

\title{{\Large {\bf Symmetries of Bianchi I space-times}}}
\author{Michael Tsamparlis and Pantelis S. Apostolopoulos \and {\it \ {\small %
Department of Physics, Section of Astrophysics-Astronomy-Mechanics, }} \and 
{\it {\small University of Athens, Zografos 15783, Athens, Greece} }}
\maketitle

\begin{abstract}
All diagonal proper Bianchi I space-times are determined which admit certain
important symmetries. It is shown that for Homotheties, Conformal motions
and Kinematic Self-Similarities the resulting space-times are defined
explicitly in terms of a set of parameters whereas Affine Collineations,
Ricci Collineations and Curvature Collineations, if they are admitted, they
determine the metric modulo certain algebraic conditions. In all cases the
symmetry vectors are explicitly computed. The physical and the geometrical
consequences of the results are discussed and a new anisitropic fluid,
physically valid solution which admits a proper conformal Killing vector, is
given.

{\bf PACS: 2.40.Hw, 4.20.-q, 4.20.Jb}
\end{abstract}

\section{Introduction}

Collineations are geometrical symmetries which are defined by the general
relation:

\begin{equation}
{\cal L}_{{\bf \xi }}{\bf \Phi }={\bf \Lambda }  \label{sx1.1}
\end{equation}
where ${\bf \Phi }$ is any of the quantities $g_{ab},\Gamma
_{bc}^{a},R_{ab},R_{bcd}^{a}$ and geometric objects constructed by them and $%
{\bf \Lambda }$ is a tensor with the same index symmetries as ${\bf \Phi }$.
Some of the well known (and most important) types of collineations are:
Conformal Killing vector (CKV) $\xi ^{a}$ defined by the requirement ${\cal L%
}_{{\bf \xi }}g_{ab}=2\psi g_{ab}$ and reducing to a Killing vector (KV)
when $\psi =0$, to a Homothetic vector field (HVF) when $\psi =$const. and
to a Special Conformal Killing vector (SCKV) when $\psi _{;ab}=0$. Proper
Affine Conformal vector (ACV) is defined by the requirement ${\cal L}_{{\bf %
\xi }}g_{ab}=2\psi g_{ab}+2H_{ab}$ where $H_{ab;c}=0,$ $\psi _{;ab}\neq 0$
and reducing to an Affine Vector (AV) when $\psi _{,a}=0$ and to a Special
Affine Conformal vector (SACV) when $\psi _{;ab}=0$. Curvature Collineation
(CC) defined by the requirement ${\cal L}_{{\bf \xi }}R_{bcd}^{a}=0$ and
finally Ricci Collineation (RC) defined by the requirement ${\cal L}_{{\bf %
\xi }}R_{ab}=0$.

Collineations other than motions (KVs) can be considered as non-Noetherian
symmetries and can be associated with constants of motion and, up to the
level of CKVs, they can be used to simplify the metric \cite{Petrov}. For
example AVs are related to conserved quantities \cite{Hojman-Nunez-Patino}
(a result used to integrate the geodesics in FRW space-times), RCs are
related to the conservation of particle number in FRW space-times \cite
{Green-Norris-Oliver-Davis} and the existence of CCs implies conservation
laws for null electromagnetic fields \cite{Katzin-Levine-Davies}.

The set of (smooth) collineations of a space-time can be related with an
inclusion relation leading to a tree like inclusion diagram \cite
{Katzin-Levine-Davies} which shows their relative hierarchy. A collineation
of a given type is proper if it does not belong to any of the subtypes in
this diagram. In order to relate a collineation to a particular conservation
law and its associated constant(s) of motion the properness of the
collineation must first be assured.

A different type of symmetry we shall discuss, which is of a kinematic
nature, is the Kinematic Self Similarity (KSS). It is defined by the
requirements \cite{Carter Henriksen 1,Carter Herinksen 2} ${\cal L}_{\xi
}u_{a}=\alpha u_{a}$, ${\cal L}_{\xi }h_{ab}=2\delta h_{ab}$ where $u^{a}$
is the four velocity of the fluid, $h_{ab}=g_{ab}+u_{a}u_{b}$ $%
(u_{a}u^{a}=-1)$ is the projection tensor normal to $u^{a}$ and $\alpha
,\delta $ are constants. A KSS reduces to a HVF when $\alpha =\delta \neq 0$
and to a KV when $\alpha =\delta =0.$ The KSS are characterised by the scale
independent ratio $\alpha /\delta $, which is known as the similarity index.
When $\alpha =0$ the KSS is of type zero (zeroth kind) and when $\delta =0$
it is of type infinite. Kinematic Self Similarity should be regarded as the
relativistic generalisation of self similarity of Newtonian Physics rather
than the generalisation of the space-time homotheties. From the physical
point of view the detailed study of cosmological models admitting KSS shows
that they can represent asymptotic states of more general models or, under
certain conditions, they are asymptotic to an exact homothetic solution \cite
{Coley-KSS,Benoit-Coley}.

A diagonal Bianchi I space-time is a spatially homogeneous space-time which
admits an abelian group of isometries $G_{3}$, acting on spacelike
hypersurfaces, generated by the spacelike KVs ${\bf \xi }_{1}=\partial _{x},%
{\bf \xi }_{2}=\partial _{y},{\bf \xi }_{3}=\partial _{z}$. In synchronous
co-ordinates the metric is:

\begin{equation}
ds^{2}=-dt^{2}+A_{\mu }^{2}(t)(dx^{\mu })^{2}  \label{sx1.2}
\end{equation}
where the metric functions $A_{1}(t),A_{2}(t),A_{3}(t)$ are functions of the
time co-ordinate only (Greek indices take the space values $1,2,3$ and Latin
indices the space-time values $0,1,2,3$). When two of the functions $A_{\mu
}(t)$ are equal (e.g. $A_{2}=A_{3}$) the Bianchi I space-times reduce to the
important class of plane symmetric space-times (a special class of the
Locally Rotational Symmetric space-times \cite{Ellis 1,Stewart and Ellis})
which admit a $G_{4}$ group of isometries acting multiply transitively on
the spacelike hypersurfaces of homogeneity generated by the vectors ${\bf %
\xi }_{1},{\bf \xi }_{2},{\bf \xi }_{3}$ and ${\bf \xi }_{4}=x^{2}\partial
_{3}-x^{3}\partial _{2}$. In this paper we are interested only in {\em %
proper diagonal} Bianchi I space-times (which in the following will be
referred for convenience simply as Bianchi I\ space-times), hence all metric
functions are assumed to be different and the dimension of the group of
isometries acting on the spacelike hypersurfaces is three.

A general Bianchi I space-time does not admit a given collineation. The
demand that it does, acts like a ''selection rule'' by selecting those
Bianchi I space-times whose metric functions $A_{\mu }(t)$ satisfy a certain
set of differential equations or algebraic conditions depending on the
collineation. These conditions do not necessarily have a solution. For
example Coley and Tupper \cite{Coley-Tupper ACVs} have determined all
space-times admitting an ACV. It is easy to check that no {\it (proper)}
Bianchi I space-time belongs to these space-times. In fact it can be shown
that the demand that a Bianchi I space-time admits an ACV leads to the
conditions $A_{3}(t)=$const. and $A_{1}(t)=A_{2}(t)$ i.e. the plane
symmetric case.

Although Bianchi I space-times are important in the study of anisotropies
and they have served as the basis for this study, it appears that their
collineations have not been considered in the literature. For example even
at the level of proper conformal symmetries, the CKV found by Maartens and
Mellin \cite{Maartens-Mellin} is a CKV in an LRS space-time and not in a
Bianchi I space-time. Perhaps this is due to the fact that the direct
solution of the collineation equations is difficult. However there are many
general results due to Hall and his co-workers - which will be referred
subsequently as they are required - which make possible the determination of
the collineations and the corresponding Bianchi I space-times that admit
them without solving any difficult differential equations.

In Section II we determine all Bianchi I space-times which admit (proper or
not) CKVs. We show that the only Bianchi I space-times which admit a proper
HVF are the Kasner-type space-times. In Section III we study the smooth RCs
and show that, provided that the Ricci tensor is non-degenerate, there are
four families of (proper) Bianchi I space-times admitting smooth RCs. The
metrics of these families are determined up to a set of algebraic conditions
among the metric functions whereas the corresponding collineation vectors
are computed in terms of a set of constant parameters. In Section IV we
study the smooth CCs of Bianchi I space-times and show that, assuming the
non degeneracy of the Ricci tensor, there are no Bianchi I space-times which
admit proper CCs. In Section V we consider the KSS\ symmetry and determine
all Bianchi I space-times which admit a KSS. A particular result which
generalises the previous result concerning the HVF, is that the only Bianchi
I space-times admitting a proper KSS (of the second kind) are the
Kasner-type space-times. In Section VI we discuss the physical implications
of these results and examine the compatibility of the physical assumptions
on the type of the matter (perfect fluid, electromagnetic field etc.) with
the various types of symmetry. An apparently new Bianchi I viscous fluid
solution is found, with non zero bulk viscous stress. The model begins with
a big-bang and isotropises at late times tending to a Robertson-Walker model
(de Sitter Universe).

\section{Conformal symmetries}

To determine the Bianchi I space-times which admit CKVs we use the theorem
of Defrise-Carter \cite{Bilyalov,Geroh,Defrise-Carter} which has been
reconsidered and improved by Hall and Steele \cite{Hall1,Hall-Steele}. This
Theorem concerns the reduction of the conformal algebra ${\cal G}$ of a
metric, to the Killing/Homothetic algebra of a (globally defined)
conformally related metric. Hall has shown that it is not always possible to
find a conformal scaling required by Defrise-Carter, but Hall and Steele
showed that the local result can be regained if one imposes at each
space-time point the restrictions (a) space-time has the same Petrov type
and (b) the dimension of ${\cal G}$ is constant. However, if the Petrov type
is I,D,II at a point these restrictions are not necessary. By a direct
computation of the Weyl tensor of the general Bianchi I metric (\ref{sx1.1})
we find that the Petrov type is either I (or its degeneracy type D which
corresponds to the LRS\ case which we ignore in this paper) or type O
(conformally flat) (in fact all Bianchi type space-times of class A are
Petrov type I or its specialisations \cite{Ellis-MacCallum}). This means
that we have two cases to consider i.e. conformally flat and non conformally
flat space-times.

\subsection{Bianchi I space-times of Petrov type I}

It is known that the maximum dimension of the conformal algebra of a
space-times of Petrov type is I or II is four \cite{Hall1}. Hence the
Bianchi I space-time admits at most exactly one {\em proper} CKV ${\bf Y}$
(the properness is assured provided that $Y^{0}\neq 0$) and consequently the
four dimensional conformal algebra ${\cal G}_{4}=\left\{ {\bf \xi }_{1},{\bf %
\xi }_{2},{\bf \xi }_{3},{\bf Y}\right\} $. From the Defrise-Carter theorem
it follows that there exists a smooth function $U(x^{a})$ such that ${\cal G}%
_{4}$ restricts to a Lie algebra of KVs for the metric $d\widehat{s}%
^{2}=U^{2}(x^{a})ds^{2}$. Because the vectors $\left\{ {\bf \xi }_{1},{\bf %
\xi }_{2},{\bf \xi }_{3}\right\} $ are KVs for both metrics we deduce that $%
U(x^{a})=U(t)$ and the type of space-time (i.e. Bianchi I) is retained.
Hence the problem of determining the CKVs of Bianchi I space-times is
reduced to the determination of the extra KV.

Assuming ${\bf Y=Y}^{\tau }(x^{i})\partial _{\tau }+Y^{\mu }(x^{i})\partial
_{\mu }$ where $\tau =\int U(t)dt$ Jacobi identities and Killing equations
imply the relations:

\begin{equation}
{\bf Y=}\partial _\tau +ax\partial _x+by\partial _y+cz\partial _z
\label{sx2.1}
\end{equation}

\begin{equation}
\hat{A}_{1}(\tau )=U(\tau )A_{1}(\tau )=e^{-a\tau },\qquad \hat{A}_{2}(\tau
)=U(\tau )A_{2}(\tau )=e^{-b\tau }\qquad \hat{A}_{3}(\tau )=U(\tau
)A_{3}(\tau )=e^{-c\tau }  \label{sx2.2}
\end{equation}
where $a\neq b\neq c$ are integration constants such that at least two are
non-zero (otherwise space-time reduces to an LRS space-time). The
commutators of the extra KV ${\bf Y}$ with the standard KVs ${\bf \xi }_{\mu
}$ are: 
\[
\lbrack {\bf \xi }_{1},{\bf Y}]=a{\bf \xi }_{1},[{\bf \xi }_{2},{\bf Y}]=b%
{\bf \xi }_{2},[{\bf \xi }_{3},{\bf Y}]=c{\bf \xi }_{3}. 
\]
In conclusion we have the following result:

{\it All Bianchi I metrics which admit a CKV are }$(a\neq b\neq c)${\it : }

\begin{equation}
ds^{2}=-dt^{2}+A_{1}^{2}(t)\left[
dx^{2}+e^{2(a-b)L(t)}dy^{2}+e^{2(a-c)L(t)}dz^{2}\right]  \label{sx2.3}
\end{equation}
{\it where: }

\begin{equation}
A_{1}(t)=\frac{1}{U(t)}e^{-a\int U(t)dt}.  \label{sx2.4}
\end{equation}
{\it The CKV is given by: }

\begin{equation}
{\bf Y}=\frac{1}{U(t)}\partial _{t}+ax\partial _{x}+by\partial
_{y}+cz\partial _{z}  \label{sx2.5}
\end{equation}
{\it and has conformal factor: }

\begin{equation}
\phi ({\bf Y)=}a+\frac{1}{U(t)}\left[ \ln \left| A_{1}(t)\right| \right]
_{,t}.  \label{sx2.6}
\end{equation}
In terms of the time coordinate $\tau $ this metric is:

\begin{equation}
ds^{2}=\frac{1}{U^{2}(\tau )}\left[ -d\tau ^{2}+e^{-2a\tau
}dx^{2}+e^{-2b\tau }dy^{2}+e^{-2c\tau }dz^{2}\right]  \label{sx2.7}
\end{equation}
It is now easy to determine all Bianchi I space-times which admit (one)
proper HVF. Indeed setting $\phi ({\bf Y)=}$const. ($\neq 0$) and using (\ref
{sx2.6}) we find $U(t)=\frac{1}{\phi t}$ from which it follows \cite
{Footnote1} (ignoring some unimportant integration constants):

\begin{proposition}
{\it The only Bianchi I space-times which admit proper HVF are the
Kasner-type space-times given by }$(a\neq b\neq c)$:
\end{proposition}

\begin{equation}
ds^{2}=-dt^{2}+t^{2\frac{\phi -a}{\phi }}dx^{2}+t^{2\frac{\phi -b}{\phi }%
}dy^{2}+t^{2\frac{\phi -c}{\phi }}dz^{2}.  \label{sx2.9}
\end{equation}
{\it The HVF is: }

\begin{equation}
{\bf Y=}\phi t\partial _t+ax\partial _x+by\partial _y+cz\partial _z
\label{sx2.10}
\end{equation}
{\it and has homothetic factor }$\phi ${\it . }

We remark that the same result can be recovered from the reduction of the
proper RCs which will be determined in Section III.

\subsection{Bianchi I space-times of Petrov type O}

\noindent The necessary and sufficient condition for conformal flatness is
the vanishing of the Weyl tensor $C_{abcd}$. By solving directly the
equations $C_{abcd}=0$ it can be shown that there exist only two families of
conformally flat Bianchi I\ metrics (we ignore the case of the
Friedmann-Robertson-Walker space-time) given by:

\begin{equation}
ds_1^2=A_3^2(\tau )ds_{RT}^2  \label{sx2.11}
\end{equation}

\begin{equation}
ds_2^2=A_3^2(\tau )ds_{ART}^2.  \label{sx2.12}
\end{equation}
The metrics $ds_{RT}^2,ds_{ART}^2$ have been found previously by Rebou\c{c}%
as-Tiomno \cite{RT} and Rebou\c{c}as-Teixeira \cite{ART} respectively and
are 1+3 (globally) decomposable space-times whose 3-spaces are spaces of
constant curvature. In synchronous co-ordinates they are:

\begin{equation}
ds_{RT}^2=dz^2-d\tau ^2+\cos ^2(\frac \tau a)dy^2+\sin ^2(\frac \tau a)dx^2
\label{sx2.13}
\end{equation}

\begin{equation}
ds_{ART}^{2}=dz^{2}-d\tau ^{2}+\cosh ^{2}(\frac{\tau }{a})dy^{2}+\sinh ^{2}(%
\frac{\tau }{a})dx^{2}  \label{sx2.14}
\end{equation}
where $d\tau =\frac{dt}{A_{3}(t)}$. Each metric admits 15 CKVs which can be
determined using standard techniques \cite{Tsamparlis-Nikolop-Apost}. In
concise notation these vectors are (we ignore the KVs $\xi _{\mu }$ which
constitute the $G_{3}$) ($k=0,1,2,3$ and $\alpha =0,1,2$):

{\bf KVs}

\[
{\bf \xi }_{k+4}=\left\{ s_{\pm }(y,a)\left[ \delta
_{k}^{0}c_{+}(x,a)+\delta _{k}^{1}s_{+}(x,a)\right] +c_{\pm }(y,a)\left[
\delta _{k}^{2}c_{+}(x,a)+\delta _{k}^{3}s_{+}(x,a)\right] \right\} \partial
_{\tau }+ 
\]

\[
+\frac{s_{\mp }(\tau ,a)}{c_{\mp }(\tau ,a)}\left\{ c_{\pm }(y,a)\left[
\delta _{k}^{0}c_{+}(x,a)+\delta _{k}^{1}s_{+}(x,a)\right] \pm s_{\pm
}(y,a)\left[ \delta _{k}^{2}c_{+}(x,a)+\delta _{k}^{3}s_{+}(x,a)\right]
\right\} \partial _{y}- 
\]

\begin{equation}
-\frac{c_{\mp }(\tau ,a)}{s_{\mp }(\tau ,a)}\left\{ s_{\pm }(y,a)\left[
\delta _{k}^{0}s_{+}(x,a)+\delta _{k}^{1}c_{+}(x,a)\right] +c_{\pm
}(y,a)\left[ \delta _{k}^{2}s_{+}(x,a)+\delta _{k}^{3}c_{+}(x,a)\right]
\right\} \partial _{x}  \label{sx2.16}
\end{equation}
\medskip

{\bf CKVs}

\begin{equation}
X_{(k)\alpha }=\pm a^{2}B_{k,\alpha }\qquad \qquad X_{(k)3}=\mp a^{2}B_{k,3}
\label{sx2.17}
\end{equation}

\begin{equation}
\phi ({\bf X}_{(k)})=B_{k}  \label{sx2.18}
\end{equation}

\begin{equation}
X_{(k+4)\alpha }=\pm a^{2}\Gamma _{k,\alpha }\qquad \qquad X_{(k+4)3}=\mp
a^{2}\Gamma _{k,3}  \label{sx2.19}
\end{equation}

\begin{equation}
\phi ({\bf X}_{(k+4)})=\Gamma _{k}  \label{sx2.20}
\end{equation}
where:

\begin{equation}
B_{k}=c_{\mp }(\tau ,a)\left\{ c_{\pm }(y,a)\left[ s_{\mp }(z,a),c_{\mp
}(z,a)\right] ,s_{\pm }(y,a)\left[ s_{\mp }(z,a),c_{\mp }(z,a)\right]
\right\}  \label{sx2.21}
\end{equation}

\begin{equation}
\Gamma _{k}=s_{\mp }(\tau ,a)\left\{ c_{+}(x,a)\left[ s_{\mp }(z,a),c_{\mp
}(z,a)\right] ,s_{+}(x,a)\left[ s_{\mp }(z,a),c_{\mp }(z,a)\right] \right\} .
\label{sx2.22}
\end{equation}
and the following conventions have been used:

1. The upper sign corresponds to RT space-time and the lower sign to ART
space-time.

2. The functions $s_{\mp }(w,a),c_{\mp }(w,a)$ are defined as follows:

\begin{equation}
(c_{+}(w,a),c_{-}(w,a))=(\cosh (\frac{w}{a}),\cos (\frac{w}{a}))
\label{sx2.24}
\end{equation}
\begin{equation}
(s_{+}(w,a),s_{-}(w,a))=(\sinh (\frac{w}{a}),\sin (\frac{w}{a}))
\label{sx2.25}
\end{equation}
The Bianchi I metrics (\ref{sx2.11}), (\ref{sx2.12}) have the same CKVs with
conformal factors $\psi ({\bf \xi }_{k+4})={\bf \xi }_{k+4}(\ln A_{3})$ and $%
\psi ({\bf X}_{A})={\bf X}_{A}(\ln A_{3})+\phi ({\bf X}_{A})$ ($A=1,2,...,8$%
). It follows that conformally flat Bianchi I space-times do not admit HVFs.
Furthermore if we enforce them to admit an extra KV they reduce to the RT
and the ART space-times which admit seven KVs.

\section{Ricci Collineations}

A RC ${\bf X=X}^a\partial _a$ is defined by the condition:

\begin{equation}
{\cal L}_{{\bf X}}R_{ab}=R_{ab,c}X^c+R_{ac}X_{,b}^c+R_{bc}X_{,a}^c=0.
\label{sx3.1}
\end{equation}
For RCs there do not exist theorems of equal power to the Theorem of
Defrise-Carter and one has to solve directly the differential equations (\ref
{sx3.1}). However there do exist some general results available which are
due to Hall and are summarised in the following statement \cite{Hall-Roy-Vaz}%
:

{\em If the Ricci tensor is of rank 4, at every point of the space-time
manifold, then the smooth (}$C^{2}$ {\em is enough) RCs form a Lie algebra
of smooth vector fields whose dimension is }$\leq 10${\em \ and }$\neq 9$%
{\em . This Lie algebra contains the proper RCs and their degeneracies. }

It has been pointed out by Hall and his co-workers that the assumption on
the order of the Ricci tensor is important. Indeed let us assume that the
order of the Ricci tensor of a Bianchi I space-time is 3. Then due to the
fact that the Ricci tensor in Bianchi I space-times is diagonal one of its
components must vanish, the $R_{11}=0$ say. This is equivalent to $R_{ab}\xi
_1^b=0$ where ${\bf \xi }_1=\partial _x$. Consider the vector field ${\bf X}%
_1=f(x^a)\partial _x$ where $f(x^a)$ is an arbitrary (but smooth) function
of its arguments. It is easy to show that ${\cal L}_{{\bf X}_1}R_{ab}=0$ so
that ${\bf X}_1$ is a Ricci collineation. Due to the arbitrariness of the
function $f(x^a)$ these Bianchi I space-times admit infinitely many smooth
RCs a result that does not help us in any useful or significant way in their
study.

In the following we consider smooth RCs and we assume that $R_{ab}$ is
non-degenerate ($\det R_{ab}\neq 0$). Equation (\ref{sx3.1}) gives the
following set of 10 differential equations (no summation over the indices $%
\mu ,\nu ,\rho =1,2,3;\mu \neq \nu \neq \rho $):

\begin{equation}
\lbrack 00]\text{ \ }R_{00,0}X^0+2R_{00}X_{,0}^0=0  \label{sx3.2}
\end{equation}

\begin{equation}
\lbrack 0\mu ]\text{ \ }R_{00}X_{,\mu }^0+R_{\mu \mu }X_{,0}^\mu =0
\label{sx3.3}
\end{equation}

\begin{equation}
\lbrack \mu \mu ]\text{ \ }R_{\mu \mu ,0}X^0+2R_{\mu \mu }X_{,\mu }^\mu =0
\label{sx3.4}
\end{equation}

\begin{equation}
\lbrack \mu \nu ]\text{ \ }R_{\mu \mu }X_{,\nu }^{\mu }+R_{\nu \nu }X_{,\mu
}^{\nu }=0  \label{sx3.5}
\end{equation}
where a comma denotes partial differentiation w.r.t. following index
co-ordinate. For convenience we set $R_{00}\equiv R_{0},R_{\mu \mu }\equiv
R_{\mu }$. Equation (\ref{sx3.2}) is solved immediately to give:

\begin{equation}
X^{0}=\frac{m(x^{\beta })}{\sqrt{|R_{0}|}}  \label{sx3.6}
\end{equation}
where $m(x^{\beta })$ is a smooth function of the spatial co-ordinates.
Using (\ref{sx3.6}) we rewrite the remaining equations (\ref{sx3.3})-(\ref
{sx3.5}) in the form:\ 

\begin{equation}
X_{,0}^\mu =-\epsilon _0\frac{\sqrt{\left| R_0\right| }}{R_\mu }m_{,\mu }
\label{sx3.7a}
\end{equation}

\begin{equation}
X_{,\mu }^{\mu }=-\frac{\left( \ln \left| R_{\mu }\right| \right) _{,0}}{2%
\sqrt{\left| R_{0}\right| }}m  \label{sx3.7b}
\end{equation}
\begin{equation}
R_{\mu }X_{,\nu }^{\mu }+R_{\nu }X_{,\mu }^{\nu }=0  \label{sx3.7c}
\end{equation}
where $\mu \neq \nu $ and $\epsilon _{0}$ is the sign of the component $%
R_{0} $.

Differentiating equation (\ref{sx3.7c}) w.r.t. $x^\rho $ we obtain:

\begin{equation}
R_\mu X_{,\nu \rho }^\mu +R_\nu X_{,\mu \rho }^\nu =0.  \label{sx3.8}
\end{equation}
Rewriting (\ref{sx3.7c}) for the indices $\mu ,\rho $, differentiating
w.r.t. $x^\nu $ and subtracting from (\ref{sx3.8}) we obtain:

\begin{equation}
R_\nu X_{,\mu \rho }^\nu -R_\rho X_{,\nu \mu }^\rho =0.  \label{sx3.9}
\end{equation}
Writing (\ref{sx3.7c}) for the indices $\nu ,\rho $, differentiating w.r.t. $%
x^\mu $ and adding to (\ref{sx3.9}) ($R_\mu \neq 0$) we get:

\begin{equation}
X_{,\mu \rho }^{\nu }=0\qquad \text{(}\mu \neq \nu \neq \rho \text{).}
\label{sx3.10}
\end{equation}
Differentiating (\ref{sx3.7a}) w.r.t. $x^{\mu }$ and (\ref{sx3.7b}) w.r.t. $%
x^{0}$ we find:

\begin{equation}
\frac{R_{\mu }}{\sqrt{\left| R_{0}\right| }}\left[ \frac{\left( \ln \left|
R_{\mu }\right| \right) _{,0}}{2\sqrt{\left| R_{0}\right| }}\right]
_{,0}=a_{\mu }  \label{sx3.11}
\end{equation}
\begin{equation}
m_{,\mu \mu }=\epsilon _{0}a_{\mu }m  \label{sx3.12}
\end{equation}
where $a_{\mu }$ are arbitrary constants.

Equations (\ref{sx3.7a})-(\ref{sx3.7c}) and (\ref{sx3.10})-(\ref{sx3.12})
constitute a set of differential equations in the variables $(R_{\mu
},X^{\mu },m)$, which can be solved in terms of $R_{0}$ and some integration
constants. These constants are constrained by a set of algebraic equations
involving the spatial components of the Ricci tensor and essentially
determine the dimension of the algebra of the RCs.

The complete set of the solutions consists of five main cases which are
summarized in TABLE I together with the corresponding algebraic constraints
and the dimension of the resulting algebra.

To find the RCs we note that in all cases the component $X^{0}$ is given by (%
\ref{sx3.6}) and it is determined in terms of the function $m(x^{\alpha }).$
Concerning the spatial components these are obtained from the following
formulas taking into consideration the fourth column of TABLE\ I.

\underline{{\bf Case A}}

\begin{equation}
X_{I}^{\mu }=c_{\mu }\left\{ -d\cdot x^{\mu }-x^{\mu }\sum_{\nu \neq \mu
}D_{\nu }x^{\nu }-\frac{D_{\mu }}{2}\left[ (x^{\mu })^{2}-\sum_{\nu \neq \mu
}\left( \frac{C_{\mu }}{C_{\nu }}\right) (x^{\nu })^{2}-\frac{\epsilon _{0}}{%
c_{\mu }^{2}R_{\mu }}\right] \right\} +\sum_{\nu \neq \mu }B_{\nu }^{\mu
}x^{\nu }  \label{sx3.13}
\end{equation}
\underline{{\bf Case B}} ($A=2,3)$%
\begin{equation}
X_{II}^{A}=c_{A}\left\{ -d\cdot x^{A}-x^{A}D_{B}x^{B}-\frac{D_{A}}{2}\left[
(x^{A})^{2}-\left( \frac{C_{A}}{C_{B}}\right) (x^{B})^{2}-\frac{\epsilon _{0}%
}{c_{A}^{2}R_{A}}\right] \right\} +\Lambda _{B}^{A}x^{B}.  \label{sx3.14}
\end{equation}
and $X_{II}^{1}=0$.

\underline{{\bf Case Ca}} 
\begin{equation}
X_{III}^{1}=-\frac{\epsilon _{0}m_{,1}}{R_{1}}\int \sqrt{\left| R_{0}\right| 
}dt+b_{2}^{1}x^{2}\quad ,\quad X_{III}^{2}=-\frac{\epsilon _{0}m_{,2}}{R_{2}}%
\int \sqrt{\left| R_{0}\right| }dt+b_{1}^{2}x^{1}  \label{sx3.15}
\end{equation}

\begin{equation}
X_{III}^{3}=-\frac{\epsilon _{0}m_{,3}}{2a\sqrt{\left| R_{0}\right| }}\left(
\ln \left| R_{3}\right| \right) _{,t}  \label{sx3.16}
\end{equation}
\underline{{\bf Case Cb}} 
\begin{equation}
X_{IV}^{1}=b_{2}^{1}x^{2}\quad ,\quad X_{IV}^{2}=b_{1}^{2}x^{1}
\label{sx3.17}
\end{equation}

\begin{equation}
X_{IV}^{3}=\frac{\epsilon _{0}D_{3}}{2c_{3}R_{3}}-c_{3}\left[ D_{3}\frac{%
(x^{3})^{2}}{2}+d\cdot x^{3}\right]  \label{sx3.18}
\end{equation}
\underline{{\bf Case D}} 
\begin{equation}
X_{V}^{\mu }=\sum_{\nu \neq \mu }b_{\nu }^{\mu }x^{\nu }+f^{\mu }(x^{0})
\label{sx3.19}
\end{equation}
where:

\begin{equation}
f^{\mu }(x^{0})=-\frac{D_{\mu }}{R_{\mu }}\int \sqrt{\left| R_{0}\right| }%
dx^{0}.  \label{sx3.20}
\end{equation}

We collect the above results in the following:

\begin{proposition}
{\it The proper smooth RCs in Bianchi I space-times can be considered in
four sets depending on the constancy of the spatial Ricci tensor components.
The first set (case A with }$R_{\mu ,0}\neq 0${\it \ for all }$\mu =1,2,3$%
{\it ) contains three families of smooth RCs consisting of either one, two
or seven RCs defined by the vector }${\bf X}_{I}${\it \ given by (\ref
{sx3.13}). The second set (case B with one }$R_{\mu ,0}=0${\it ) consists of
two families of one and four RCs defined by the vector }${\bf X}_{II}${\it \
given by (\ref{sx3.14}). The third set (case C with two }$R_{\mu ,0}=0)${\it %
\ consists of five families with three, three, three, seven and three RCs
are given by the vector fields }${\bf X}_{III},{\bf X}_{IV}${\it \ defined
in (\ref{sx3.15})-(\ref{sx3.16}), (\ref{sx3.17})-(\ref{sx3.18}). Finally the
last set (case D all }$R_{\mu ,0}=0)${\it \ contains one family of seven RCs
given by the vector field }${\bf X}_{V}${\it \ defined in (\ref{sx3.20}). In
each family of RCs the spatial Ricci tensor components are given in terms of
the time component }$R_{0}$ {\it and the coefficients of the vector fields
are constrained with the spatial components of the Ricci tensor via
algebraic conditions. }
\end{proposition}

We note that RCs do not fix the metric up to a set of constants as the CKVs
(and the lower symmetries) do but instead they impose algebraic conditions
on the metric functions. Thus in general one should expect many families of
Bianchi I space-times admitting RCs.

\section{Curvature Collineations}

Curvature collineations are necessarily Ricci Collineations and one is
possible to determine them from the results of the last section. The easiest
way to do this would appear to use algebraic computing algorithms \cite
{Melfo-Nunez} and compute directly the Lie derivative of the curvature
tensor for the RCs found in the last section. However this is not so obvious
because although we know the RC we do not know the metric functions. Hence,
in general, one expects to arrive at a system of differential equations
among the metric functions $A_\mu (t)$ whose solution will give the answer.

However a study of the relevant literature shows that there are enough
general results which allow one to determine the CCs without solving any
differential equations. The CCs have many of the pathologies of RCs. For
example for any positive integer $k$ there are CCs which are $C^k$ but not $%
C^{k+1}.$ Furthermore they may form an infinite dimensional vector space
which is not a Lie algebra under the usual Lie bracket operation. However if
one considers the $C^\infty $ CCs only (loosing in that case the ones that
are not smooth) then they do form a Lie algebra which is a subalgebra of the
(smooth) RCs algebra \cite{Hall-Da Costa I,Hall-Da Costa II}.

For the determination of CCs in Bianchi I space-times it is enough to use
the following result from an early work of Hall \cite{Hall 1983}:

{\it If the curvature components are such that at every point of a
space-time M the only solution of the equation }$R_{abcd}k^{d}=0${\it \ is }$%
k^{d}=0${\it \ then every CC on M is a HVF.}

Let us assume that in a Bianchi I\ space-time the equation $R_{abcd}k^{d}=0$%
{\it \ }has a solution{\it \ }$k^{a}\neq 0$. Then it follows that equation $%
R_{ab}k^{d}=0${\it \ }admits a non-vanishing solution which is impossible
because $R_{ab}$ is non-degenerate. Hence according to the above statement
all $C^{\infty }$ CC in Bianchi I space-times are HVFs or, equivalently {\it %
there are no Bianchi I space-times (with non-degenerate Ricci tensor) which
admit proper CCs. }

\section{Kinematic Self Similarities}

As it has been mentioned in the Introduction, Kinematic Self Similarities
are not geometric symmetries (i.e. collineations). They are kinematic
symmetries/constraints which involve the 4-velocity of the fluid (or in
empty space-times a timelike unit vector field) defined by the conditions:

\begin{equation}
{\cal L}_{X}u_{a}=\alpha u_{a}\qquad {\cal L}_{X}h_{ab}=2\delta h_{ab}
\label{sx5.1}
\end{equation}
where $h_{ab}=g_{ab}+u_{a}u_{b}$ projects normally to $u_{a}$ and $\alpha
,\delta $ are constants.

Kinematic self similarities have been studied by Sintes\cite{Sintes-KSS} who
determined all LRS perfect fluid space-times which admit a KSS. Our aim in
this Section is to determine the KSS of (proper) Bianchi I metrics without
any restriction on the type of the fluid except that we assume that the
fluid 4-velocity is orthogonal to the group orbits i.e. $u^{a}=\delta
_{0}^{a}$. This assumption enforces the commutator of a KSS ${\bf X}$ with
the three KVs ${\bf \xi }_{\mu }$\ to be a KV\cite{footnote2}. Hence we
write:

\begin{equation}
\lbrack {\bf \xi }_{\mu },{\bf X}]=X_{,\mu }^{a}\partial _{a}=a_{\mu }^{\nu }%
{\bf \xi }_{\nu }  \label{sx5.2}
\end{equation}
where $a_{\mu }^{\nu }$ are constants. Integrating we find:

\begin{equation}
X^{0}=X^{0}(t)\qquad \text{and}\qquad X^{\mu }=a_{\nu }^{\mu }x^{\nu
}+f^{\mu }(t)  \label{sx5.3}
\end{equation}
where $f^{\mu }(t)$ are arbitrary smooth functions of their argument. The
first of equations (\ref{sx5.1}) gives:

\[
X^{0}=\alpha t+\beta 
\]
where $\beta $ is an integration constant and (without loss of generality) $%
f^{\mu }(t)=$const.$=0$. The second equation of (\ref{sx5.1}) gives the
following conditions among the metric functions:

\begin{equation}
a_\nu ^\mu (A_\mu )^2+a_\mu ^\nu (A_\nu )^2=0  \label{sx5.4}
\end{equation}

\begin{equation}
a_{\mu }^{\mu }+(\alpha t+\beta )\frac{d(\ln A_{\mu })}{dt}=\delta
\label{sx5.5}
\end{equation}
where $\mu \neq \nu $ and a dot denotes differentiation w.r.t. $t$. Equation
(\ref{sx5.4}) means that, in order to avoid the plane symmetric case, we
must take $a_{\nu }^{\mu }=0$ for $\mu \neq \nu $. Therefore we have the
following conclusion ($a_{\mu }^{\mu }\equiv a_{\mu }$):

\begin{proposition}
{\it The Bianchi I space-times whose metric functions satisfy the relation (}%
$a_{\mu }\epsilon R$){\it :}
\end{proposition}

\begin{equation}
a_{\mu }+(\alpha t+\beta )\frac{d(\ln A_{\mu })}{dt}=\delta  \label{sx5.6}
\end{equation}
{\it admit the proper KSS:}

\begin{equation}
{\bf X}=(\alpha t+\beta )\partial _{t}+a_{1}x\partial _{x}+a_{2}y\partial
_{y}+a_{3}z\partial _{x}.  \label{sx5.7}
\end{equation}
In view of equation (\ref{sx5.6}) we have two distinct cases to consider,
namely $\alpha =0$ (type zero) and $\alpha \neq 0$.

\noindent \underline{Case $\alpha =0$} $(\delta \neq 0)$.

For $\beta =0$ equation (\ref{sx5.6}) is trivially satisfied i.e. {\it all
Bianchi I space-times admit the (zeroth kind) KSS:}

\begin{equation}
{\bf Z}=\delta (x\partial _x+y\partial _y+z\partial _x).  \label{sx5.8}
\end{equation}
For $\beta \neq 0$ we have:

\begin{equation}
A_{\mu }(t)=e^{\frac{\delta -a_{\mu }}{\beta }t}.  \label{sx5.9}
\end{equation}
It is easy to show that in this case space-time admits the KV:

\begin{equation}
{\bf Y}=\partial _{t}+\frac{a_{1}-\delta }{\beta }x\partial _{x}+\frac{%
a_{2}-\delta }{\beta }y\partial _{y}+\frac{a_{3}-\delta }{\beta }z\partial
_{x}  \label{sx5.10}
\end{equation}
and becomes homogeneous. (The KV (\ref{sx5.10}) can also be found from the
reduction of the results of Section II).

\underline{Case $\alpha \neq 0$ }

In this case the solution of (\ref{sx5.6}) is:

\begin{equation}
A_{\mu }(t)=(\alpha t+\beta )^{\frac{\delta -a_{\mu }}{\alpha }}
\label{sx5.11}
\end{equation}
and leads to the conclusion that:

{\it The only Bianchi I\ space-times with co-moving fluid which admit a KSS
of the second kind are the Kasner type space-times. }

We note that the Bianchi I space-times with metric functions given by
equation (\ref{sx5.11}) also admit the HVF: 
\begin{equation}
\alpha \partial _{t}+(\alpha +a_{1}-\delta )x\partial _{x}+(\alpha
+a_{2}-\delta )y\partial _{y}+(\alpha +a_{3}-\delta )z\partial _{x}
\label{sx5.12}
\end{equation}
with homothetic factor $\alpha $ (see Section II).

\section{Discussion}

Working with purely geometric methods we have succeeded to determine all
(proper and diagonal) Bianchi I space-times which admit certain (and the
most important) collineations. In many cases the explicit form of the
metrics has been given (CKVs, HVFs, KSS), in others the metrics are defined
up to a set of conditions on the metric functions (RCs) and finally it has
been shown that there are not Bianchi I space-times which admit CCs and
ACVs. In all cases the collineation vector has been determined (whenever it
exists).

In order to establish the physical significance of these general geometrical
results we address the following questions for each type of collineation:

{\em 1. Are there any Bianchi I metrics among the ones selected by one of
the collineations considered, which satisfy the energy conditions, so that
they can be used as potential space-time metrics?}

{\em 2. If there are, are the known Bianchi I solutions among these
solutions?}

In the following we take the cosmological constant $\Lambda =0$.

\subsection{The case of CKVs}

Of interest is only the non-conformally flat metrics (\ref{sx2.4}) (or (\ref
{sx2.5})) which admit the CKV (\ref{sx2.7}) with conformal factor (\ref
{sx2.6}).

One general result is that the CKV is inheriting, that is, the fluid flow
lines are preserved under Lie transport along the CKV\cite{Coley-Tupper
Inheriting CKV}.

Concerning the dynamical results we consider three main cases:\ perfect
fluids, non null Einstein-Maxwell solutions and imperfect fluid solutions
(the case of null Einstein-Maxwell field\cite{footnote4} is excluded because
the Segr\'{e} type of Bianchi I space-times is $[1,111]$ or degeneracies of
this type).

\underline{Perfect Fluid Solutions}

In this case the Segr\'{e} type of space-time is $[1,(111)]$ and all spatial
eigenvalues $\lambda _{\mu }$ are equal. Moreover $\lambda _{\mu }=G_{\mu
}^{\mu }$ where $G_{ab}$ is the Einstein tensor. Considering $\lambda
_{2}=\lambda _{3}$ and using (\ref{sx2.6}) we find $U=\frac{1}{Bt}$ where $B=%
\frac{a+b+c}{2}$. Finally replacing $U(t)$ in the line element (\ref{sx2.4})
we find that the resulting space-time is a Kasner type space-time and the
CKV reduces to a HVF in agreement with the general result that {\it %
orthogonal spatially homogeneous perfect fluid space-times do not admit any
inheriting proper CKV.}\cite{Coley-Tupper Inheriting CKV}{\it \ }Most of the
known Bianchi I solutions concern perfect fluid solutions\cite{Jacobs }.
None of these solutions admit a CKV.

\noindent \underline{Einstein Maxwell solutions}

For these fields the Segr\'{e} type of the Einstein tensor is $[(1,1)(11)]$
hence $\lambda _{2}=\lambda _{3}$ and $\lambda _{0}=\lambda _{1}$. The first
equality implies again that the metric reduces to a Kasner type metric and
in fact to a vacuum solution (because if we force a Kasner type metric to
represent a (necessarily non null) Einstein-Maxwell field it reduces to a
vacuum solution).

\begin{proposition}
There do not exist Bianchi I (non-null) Einstein-Maxwell space-times which
admit a CKV or a HVF.
\end{proposition}

The two Bianchi I solutions with electromagnetic field found by Datta \cite
{Datta } and by Rosen \cite{Rosen } do not admit a CKV or a HVF.

\noindent \underline{Anisotropic fluid solutions}

The above results indicate that the Bianchi I\ metrics (\ref{sx2.4}) can
represent only anisotropic fluid space-times. Recently anisotropic fluid
Bianchi I cosmological models have been investigated extensively using a
dynamical system approach and the truncated Israel-Stewart theory of
irreversible thermodynamics. It has been found that in these models,
anisotropic stress leads to models which violate the weak energy condition,
thus they are unphysical or they lead to the creation of a periodic orbit 
\cite
{Burd-Coley,Coley-Hoogen1,Maartens,Coley-Hoogen2,Coley-Hoogen3,Coley-Hoogen-Maartens}%
.

In order to find one such solution which will be physically viable we
consider the following two restrictions:

1. $c=0$ and

2. The algebraic type of matter (equivalently Einstein) tensor is $%
[(1,1)11]. $

Setting $\lambda _{0}=\lambda _{3}$ we obtain the condition:

\begin{equation}
2(\ln M)_{,\tau \tau }-2\left[ (\ln M)_{,\tau }\right] ^2+b^2+a^2=0
\label{sx6.13}
\end{equation}
whose solution is:

\begin{equation}
M(\tau )=\frac{1}{\sinh k\tau }\text{ },\text{ }\frac{1}{\cosh k\tau }
\label{sx6.14}
\end{equation}
where $k^{2}=\frac{a^{2}+b^{2}}{2}$. We keep the solution $M(\tau )=\frac{1}{%
\sinh k\tau }$ because the other violates all energy conditions. This gives $%
U(\tau )=\sinh k\tau $ or $U(t)=\sinh ^{-1}kt$ and finally we obtain the
metric: 
\begin{equation}
ds^{2}=-dt^{2}+\sinh ^{2\frac{k-a}{k}}\frac{kt}{2}\cosh ^{2\frac{k+a}{k}}%
\frac{kt}{2}dx^{2}+\sinh ^{2\frac{k-b}{k}}\frac{kt}{2}\cosh ^{2\frac{k+b}{k}}%
\frac{kt}{2}dy^{2}+\sinh ^{2}ktdz^{2}.  \label{sx6.15}
\end{equation}
This new Bianchi I space-time describes a viscous fluid and satisfies the
weak and the dominant energy conditions (a description of these energy
conditions in terms of the eigenvalues of the stress-energy tensor is given
in the Appendix) provided $ab>0$ and $ka<0$. The strong energy condition is
violated. It also admits the proper CKV ${\bf X}=\sinh kt\partial
_{t}+ax\partial _{x}+by\partial _{y}$ with conformal factor $\phi ({\bf X}%
)=k\cosh kt$.

To study the physics of the new solution we consider the stress-energy
tensor $T_{ab}$ and using the standard Eckart theory we write:

\begin{equation}
T_{ab}=\mu u_{a}u_{b}+(\bar{p}-\zeta \theta )h_{ab}-2\eta \sigma _{ab}
\label{sx6.16}
\end{equation}
where $\zeta ,\eta \geq 0$ are the bulk and the shear viscosity coefficients
and $\bar{p}$ is the isotropic pressure in the absence of dissipate
processes i.e. $\zeta =0$ (equilibrium state). It is easy to show that
vanishing of $\zeta $ together with a linear barotropic equation of state $%
\bar{p}=(\gamma -1)\mu $ (where $\gamma \in [1,2]$) lead to the condition $%
a=-b,$ which violates the weak energy condition ($ab<0$). Thus we restrict
our study to the case where $\zeta \neq 0$ which is of cosmological
interest. For example inflation driven by a viscous fluid necessarily
involves bulk viscous stress\cite{Maartens-Mendez}. Moreover cosmological
models which include viscosity can be used in an attempt to interpret the
observed highly isotropic matter distribution\cite{Coley-GRG}. In fact it
has been shown that viscosity plays a significant role in the isotropisation
of the cosmological models.\cite{Belinski-Khalatnikov}

Using standard methods we find for the kinematic and the dynamic variables
of the model:

\begin{eqnarray}
\mu &=&\frac{3k^{2}\cosh ^{2}kt-2k(b+a)\cosh kt+ab}{\sinh ^{2}kt}  \nonumber
\\
&&  \nonumber \\
\zeta &=&\frac{1}{\theta }\left[ \mu +\bar{p}+\frac{2k(a+b)\cosh
kt-2(k^{2}+ab)}{3\sinh ^{2}kt}\right]  \label{sx6.17} \\
&&  \nonumber \\
\theta &=&\frac{3k\cosh kt-(b+a)}{\sinh kt}  \nonumber
\end{eqnarray}

\begin{eqnarray}
\sigma _{11} &=&\frac{(b-2a)}{6}\sinh ^{\frac{k-2a}{k}}\frac{kt}{2}\cosh ^{%
\frac{k+2a}{k}}\frac{kt}{2}  \nonumber \\
&&  \nonumber \\
\sigma _{22} &=&\frac{(a-2b)}{6}\sinh ^{\frac{k-2b}{k}}\frac{kt}{2}\cosh ^{%
\frac{k+2b}{k}}\frac{kt}{2}  \label{sx6.18} \\
&&  \nonumber \\
\sigma _{33} &=&\frac{(a+b)}{6}\sinh kt  \nonumber
\end{eqnarray}
\begin{equation}
\sigma ^{2}=\frac{a^{2}-ab+b^{2}}{3\sinh ^{2}kt}  \label{sx6.18a}
\end{equation}
where $2\sigma ^{2}=\sigma _{ab}\sigma ^{ab}$ and:

\begin{equation}
\eta =\frac{a+b}{2\sinh kt}-k\coth kt.  \label{sx6.19}
\end{equation}
The explicit computation of the bulk viscosity $\zeta $ requires the
adoption of a specific equation of state in order to guarantee that $\zeta $
is positive definite. However the simple choice $a+b>0$ and $k<0$ ensures
that $\zeta \geq 0$ (since $\theta <0$) and $\eta >0$ provided that $\bar{p}%
\leq p_{eff}\equiv \bar{p}-\zeta \theta $.

Concerning the asymptotic behavior of the model we have $\lim\limits_{t%
\rightarrow \infty }\sigma =0$ provided that $k<0$ and the model isotropizes
at late times. Furthermore $\lim\limits_{t\rightarrow \infty }\theta =-3k$ , 
$\lim\limits_{t\rightarrow \infty }R=12k^{2},$ $(\mu +p_{eff})_{t\rightarrow
\infty }=0$ hence the model corresponds to the flat FRW space-time i.e. the
de Sitter universe. This also follows directly from the metric (\ref{sx6.15}%
) if we consider the limit $t\rightarrow \infty $. In this limit the CKV\ $%
{\bf X}$ degenerates to a KV. In addition using (\ref{sx6.17}) it can be
shown that there is a cosmological singularity of Kasner type at a {\em %
finite} time in the past i.e. there is a $t=t_{0}$ at which the energy
density vanishes, whereas the initial singularity occurs at $t=0$.

\subsection{The case of Ricci Collineations}

There are four families of Bianchi I\ space-times which admit RCs and
furthermore the metric in these families is fixed only up to a set of
algebraic conditions. Due to this generality we are obliged to consider
again special cases and the best choice is perfect fluid solutions.

The algebraic type of the Ricci tensor for a perfect fluid is $[1,(111)]$
which implies the condition:

\begin{equation}
\frac{R_{11}}{A_{1}^{2}}=\frac{R_{22}}{A_{2}^{2}}=\frac{R_{33}}{A_{3}^{2}}.
\label{sx6.30}
\end{equation}
This immediately excludes the last three families (B,Ca,Cb,D) of TABLE I and
we are left with family A only. From the first column of TABLE\ I we read $%
R_{\mu }=C_{\mu }e^{\int 2c_{\mu }\sqrt{\left| R_{0}\right| }dt}$ hence (\ref
{sx6.30}) imply:

\begin{equation}
\frac{C_{1}}{C_{2}}e^{2(c_{1}-c_{2})\int \sqrt{\left| R_{0}\right| }dt}=%
\frac{A_{1}^{2}}{A_{2}^{2}}\qquad \text{and}\qquad \frac{C_{1}}{C_{3}}%
e^{2(c_{1}-c_{3})\int \sqrt{\left| R_{0}\right| }dt}=\frac{A_{1}^{2}}{%
A_{3}^{2}}.  \label{sx6.31}
\end{equation}
In order to avoid the FRW metric ($A_{1}\propto A_{2}\propto A_{3})$ and the
plane symmetric metric (e.g. $A_{1}\propto A_{2}$) we demand $c_{1}\neq
c_{2}\neq c_{3}$. Then from the third column of TABLE\ I we have that only
case A$_{2}$ survives and furthermore $D_{\mu }=b_{\nu }^{\mu }=0$. Setting
the constant $d=1$ we find from (\ref{sx3.13}):

\begin{equation}
{\bf X}=\frac{1}{\sqrt{R_{00}}}\partial _{0}-c_{1}x^{1}\partial
_{1}-c_{2}x^{2}\partial _{2}-c_{3}x^{3}\partial _{3}.  \label{sx6.32}
\end{equation}
We conclude that all perfect fluid Bianchi I space-times which satisfy (\ref
{sx6.31}) admit a RC of the form (\ref{sx6.32}).

As far as we aware all existing perfect fluid solutions in Bianchi I
space-times concern perfect fluids with a linear barotropic equation of
state $p=(\gamma -1)\mu $ ($\gamma \in [1,2]$). In order to compare these
solutions with the above perfect fluid solutions we assume that the later
also satisfy a linear barotropic equation of state. Then from the field
equations we obtain:

\begin{equation}
R_{ab}=\mu \left[ \frac{(2-\gamma )}{2}g_{ab}+\gamma u_{a}u_{b}\right] .
\label{sx6.33}
\end{equation}
(The value $\gamma =2$ is excluded because then the Ricci tensor becomes
degenerated). The 00 conservation equation gives:\ 

\begin{equation}
\mu _{,t}+\gamma \mu \theta =0  \label{sx6.34}
\end{equation}
where $\theta =u_{;a}^{a}$ is the expansion of the fluid. Introducing the
scale factor or ''mean radius'' $S^{3}=A_{1}A_{2}A_{3}$ we find $\theta
=\left( \ln S^{3}\right) _{,t}$ and the solution of equation (\ref{sx6.34})
is:

\begin{equation}
\mu =\frac{M}{S^{3\gamma }}  \label{sx6.35}
\end{equation}
where $M=$ constant. Using (\ref{sx6.31}), (\ref{sx6.33}) and (\ref{sx6.35})
we find: 
\begin{equation}
\mu =\frac{\text{const.}}{t^{2}}.  \label{sx6.36}
\end{equation}
Combining this with (\ref{sx6.30}) and (\ref{sx6.31}) we find $%
A_{1}\varpropto t^{p},A_{2}\varpropto t^{q},A_{3}\varpropto t^{r}$ i.e. the
resulting Bianchi space-time is a Kasner type space-time which is a
contradiction because a Kasner type space-time with a perfect fluid leads to
stiff matter\cite{Jacobs } i.e. $\gamma =2$. Thus we have proved:

\begin{proposition}
{\it All perfect fluid Bianchi I space-times whose metric functions satisfy
condition (\ref{sx6.31}) admit a RC of the form (\ref{sx6.32}). In addition
perfect fluid Bianchi I space-times with linear barotropic equation of state
do not admit proper RCs. }
\end{proposition}

Finally it should be pointed that although spatially homogeneous perfect
fluid space-times must satisfy a barotropic equation of state $p=p(\mu )$
this equation need not necessarily be linear. However it was proved recently
that the asymptotic behaviour of such models is similar to the case of a
linear barotropic equation of state \cite{Rendall}.

{\LARGE Acknowledgments}

We thank the referee for useful and accurate remarks and suggestions. One of
the authors (P.S.A.) was partially supported by the Hellenic Fellowship
Foundation (I.K.Y.). \bigskip

{\LARGE Appendix}

It is well known that Bianchi I space-times have zero heat flux and the
4-velocity $u^{a}$ is an eigenvector of the energy momentum tensor.
Consequently the only possible algebraic Segr\'{e} type of their energy
momentum tensor is $[1,111]$ and its degeneracies \cite{Hall Arabic
Journal,Hall-Int.J.Phys.,Footnote5}. Furthermore, for this type of energy
momentum tensors it has been shown that the energy conditions take the
following form \cite{Wald,Kollasis-Santos-Tsoubelis}:

\underline{{\em Weak energy condition}}

\begin{equation}
-\lambda _{0}\geqslant 0\text{ and }-\lambda _{0}+\lambda _{\alpha
}\geqslant 0  \tag{A1}
\end{equation}

\underline{{\em Dominant energy condition}} 
\begin{equation}
-\lambda _{0}\geqslant 0\text{ and }\left| \lambda _{\alpha }\right|
\leqslant -\lambda _{0}  \tag{A2}
\end{equation}

\underline{{\em Strong energy condition}}

\begin{equation}
-\lambda _{0}\geqslant 0\text{ and }-\lambda _{0}+\sum\limits_{\alpha
}\lambda _{\alpha }\geqslant 0  \tag{A3}
\end{equation}
where $\lambda _{0}$ is the eigenvalue of the timelike eigenvector and $%
\lambda _{\alpha }$ are the eigenvalues of the spacelike eigenvectors.

{\bf TABLE I}: The table contains the complete set of solutions of the Ricci
Collineation equations. The solutions are specified in terms of $R_{0}$ and
some integration constants together with the algebraic constraints (if any)
which $R_{\mu }$ must satisfy. The second column contains the spatial
components of the Ricci tensor, the third the function $m(x^{\alpha })$, the
fourth column the constraints on the integration constants and the last
column the number of proper RCs. The indices $\mu ,\nu ,\rho =1,2,3$ , $\mu
\neq \nu \neq \rho $ , $A,B=2,3$ $A\neq B$ and there is no summation over
repeated indices. The functions $c_{\varepsilon }(x^{3},a),s_{\varepsilon
}(x^{3},a)$ are given in (23) and (24).

\begin{center}
\begin{tabular}{ccccc}
\hline\hline
{\bf Case} & $R_{\mu }$ & $m(x^{\alpha })$ & {\bf Constraints} & 
\begin{tabular}{l}
{\bf No of} \\ 
{\bf proper} \\ 
{\bf RCs}
\end{tabular}
\\ \hline
{\bf A}$_{1}$ & $R_{\mu }B_{\nu }^{\mu }+R_{\nu }B_{\mu }^{\nu }=0$ & $0$ & 
none & $3$ \\ \hline
{\bf A}$_{2}$ & $R_{\mu }=C_{\mu }e^{\int 2c_{\mu }\sqrt{\left| R_{0}\right| 
}dx^{0}}$ & $d$ & $
\begin{tabular}{l}
$c_{\mu }\neq c_{\nu }$ \\ 
$D_{\mu }=B_{\mu }^{\nu }=0$%
\end{tabular}
$ & $1$ \\ \hline
{\bf A}$_{3}$ & 
\begin{tabular}{l}
$R_{\mu }=C_{\mu }e^{\int 2c_{\mu }\sqrt{\left| R_{0}\right| }dx^{0}}$ \\ 
$R_{\rho }B_{\nu }^{\rho }+R_{\nu }B_{\rho }^{\nu }=0$%
\end{tabular}
& $d$ & $
\begin{tabular}{l}
$c_{\rho }=c_{\nu }\neq c_{\mu }$ \\ 
$D_{\mu }=B_{\mu }^{\nu }=0$%
\end{tabular}
$ & $2$ \\ \hline
{\bf A}$_{4}$ & 
\begin{tabular}{l}
$R_{\mu }=C_{\mu }e^{\int 2c_{\mu }\sqrt{\left| R_{0}\right| }dx^{0}}$ \\ 
$R_{\rho }B_{\nu }^{\rho }+R_{\nu }B_{\rho }^{\nu }=0$%
\end{tabular}
& $\dsum\limits_{\mu =1,2,3}D_{\mu }x^{\mu }+d$ & $c_{\mu }=c_{\nu }$ & $7$
\\ \hline
{\bf B}$_{1}$ & 
\begin{tabular}{l}
$R_{1,0}=0$ \\ 
$R_{A}=C_{A}e^{\int 2c_{A}\sqrt{\left| R_{0}\right| }dx^{0}}$%
\end{tabular}
& $d$ & $
\begin{tabular}{l}
$c_{A}\neq c_{B}$ \\ 
$D_{A}=\Lambda _{A}^{B}=0$%
\end{tabular}
$ & $1$ \\ \hline
{\bf B}$_{2}$ & 
\begin{tabular}{l}
$R_{1,0}=0$ \\ 
$R_{A}=C_{A}e^{\int 2c_{A}\sqrt{\left| R_{0}\right| }dx^{0}}$ \\ 
$R_{A}\Lambda _{B}^{A}+R_{B}\Lambda _{A}^{B}=0$%
\end{tabular}
& $\dsum\limits_{A=2,3}D_{A}x^{A}+d$ & $c_{A}=c_{B}$ & $4$ \\ \hline
{\bf Ca}$_{1}$ & 
\begin{tabular}{l}
$R_{1,0}=R_{2,0}=0$ \\ 
$R_{1}b_{2}^{1}+R_{2}b_{1}^{2}=0$ \\ 
$R_{3}=\epsilon _{3}C_{3}^{2}\cosh ^{2}\left[ \frac{(\epsilon
_{3}a)^{1/2}\int \sqrt{\left| R_{0}\right| }dt}{C_{3}}\right] $%
\end{tabular}
& 
\begin{tabular}{l}
$\beta _{1}s_{\varepsilon }\left( x^{3},\frac{1}{\sqrt{\left| \epsilon
_{0}a\right| }}\right) +$ \\ 
$\beta _{2}c_{\varepsilon }\left( x^{3},\frac{1}{\sqrt{\left| \epsilon
_{0}a\right| }}\right) $%
\end{tabular}
& $
\begin{tabular}{l}
$\epsilon _{3}a>0$ \\ 
$\varepsilon =sign(\epsilon _{0}a)$%
\end{tabular}
$ & $3$ \\ \hline
{\bf Ca}$_{2}$ & 
\begin{tabular}{l}
$R_{1},R_{2}$ same as {\bf Ca}$_{1}$ \\ 
$R_{3}=\epsilon _{3}C_{3}^{2}\sinh ^{2}\left[ \frac{\left| (\epsilon
_{3}a)\right| ^{1/2}\int \sqrt{\left| R_{0}\right| }dt}{C_{3}}\right] $%
\end{tabular}
& same as {\bf Ca}$_{1}$ & $\epsilon _{3}a<0$ & $3$ \\ \hline
{\bf Ca}$_{3}$ & 
\begin{tabular}{l}
$R_{1},R_{2}$ same as {\bf Ca}$_{1}$ \\ 
$R_{3}=\epsilon _{3}C_{3}^{2}\cos ^{2}\left[ \frac{\left| (\epsilon
_{3}a)\right| ^{1/2}\int \sqrt{\left| R_{0}\right| }dt}{C_{3}}\right] $%
\end{tabular}
& same as {\bf Ca}$_{1}$ & $\epsilon _{3}a<0$ & $3$ \\ \hline
{\bf Ca}$_{4}$ & 
\begin{tabular}{l}
$R_{1},R_{2}$ same as {\bf Ca}$_{1}$ \\ 
$R_{3}=-\epsilon _{3}a\left( \int \sqrt{\left| R_{0}\right| }dt\right) ^{2}$%
\end{tabular}
& 
\begin{tabular}{l}
$\sum\limits_{\alpha }m_{\alpha }(x^{3})x^{\alpha }$ \\ 
$m_{\alpha }=\gamma _{\alpha }^{1}s_{\varepsilon }\sqrt{\left| \epsilon
_{0}a\right| }x^{3}+$ \\ 
$+\gamma _{\alpha }^{2}c_{\varepsilon }\sqrt{\left| \epsilon _{0}a\right| }%
x^{3}$%
\end{tabular}
& $\varepsilon =sign(\epsilon _{0}a)$ & $7$ \\ \hline
{\bf Cb} & 
\begin{tabular}{l}
$R_{1},R_{2}$ same as {\bf Ca}$_{1}$ \\ 
$R_{3}=C_{3}e^{\int 2c_{3}\sqrt{\left| R_{0}\right| }dt}$%
\end{tabular}
& $D_{3}x^{3}+d$ & $c_{3}\neq 0$ & $3$ \\ \hline
{\bf D} & 
\begin{tabular}{l}
$R_{1,0}=R_{2,0}=R_{3,0}=0$ \\ 
$R_{\mu }B_{\nu }^{\mu }+R_{\nu }B_{\mu }^{\nu }=0$%
\end{tabular}
& $\dsum\limits_{\mu =1,2,3}D_{\mu }x^{\mu }+d$ & none & $7$ \\ \hline\hline
\end{tabular}
\end{center}


\begin{thebibliography}{99}
\bibitem{footnote4}  We recall that the Segr\'{e} type of a null
Einstein-Maxwell field is $[(2,11)]$.
\bibitem{Footnote5}  The types $[3,1]$ and $[z\bar{z},11]$ are not allowed
from the energy conditions and will not be considered further.
\end{thebibliography}
\end{document}